\documentclass{llncs}
\usepackage{makeidx}  
\begin{document}
\frontmatter          
\pagestyle{headings}  
\addtocmark{Evaluation of freely available open source software for clinical concept recognition} 
\mainmatter              
\title{Evaluation of YTEX and MetaMap for clinical concept recognition}
\titlerunning{YTEX and MetaMap Evaluation}  
%
\author{John David Osborne, Binod Gyawali \and Thamar Solorio}
\authorrunning{John David Osborne et al.} 
%
\tocauthor{John David Osborne, Binod Gyawali and Thamar Solorio}
\institute{University of Alabama at Birmingham, Birmingham AL 35294, USA, \\
\email{\{ozborn,gyawali,solorio\}@uab.edu},\\  WWW home page: \texttt{http://coral.cis.uab.edu/}
}

\maketitle              

\begin{abstract}
We used MetaMap and YTEX as a basis for the construction of two separate systems to participate in the 2013 ShARe/CLEF eHealth Task 1\cite{Suominen:h}, the recognition of clinical concepts. No modifications were directly made to these systems, but output concepts were filtered using stop concepts, stop concept text and UMLS semantic type. Concept boundaries were also adjusted using a small collection of rules to increase precision on the strict task. Overall MetaMap had better performance than YTEX on the strict task, primarily due to a 20\% performance improvement in precision. In the relaxed task YTEX had better performance in both precision and recall giving it an overall F-Score 4.6\% higher than MetaMap on the test data. Our results also indicated a 1.3\% higher accuracy for YTEX in UMLS CUI mapping.

\end{abstract}
\section{Introduction}

The mapping of natural language text to a defined dictionary of concepts is one of the signature tasks of information extraction and a variety of software has been written to address this problem. We informally assessed a small set of open source and freely available software for this task, before selecting MetaMap \cite{Aronson:2010kv} and YTEX \cite{garla:2011ml} for a more formal evaluation. Our objective was to find "off the shelf" software that could find disease mentions in clinical text and be robust with regard to clinical domain as well as document structure and syntax. Although MetaMap was designed for publications, it continues to be on clinical text, often as a standard of comparison. Although YTEX has received comparatively little attention, it is based on the more popular cTAKES \cite{Savova:2010hy} system developed for the Mayo Clinic and is specifically designed for clinical text. We did download and informally evaluate cTAKES but found the dictionary lookup annotator component by matching tokenized Lucene (http://lucene.apache.org/) indexed dictionary entries and was therefore unable to distinguish different concepts sharing the same lexical tokens. Recently YTEX has improved on cTAKES dictionary lookup by adding a sense disambiguation component \cite{garla2012knowledge} that allows the most appropriate concept for that text. It uses the adapted Lesk Method \cite{banerjee2002adapted} to compute semantic similarly over a context window, whereas MetaMap uses a series of weighted heuristics to select the appropriate candidate \cite{Aronson:2001jv}. As part of our evaluation, we hoped to learn the relative strengths and weakness of these two very different approaches.

\section{Methodology}

The CORAL.1 system utilizes the 2012 version of MetaMap for concept recognition whereas the CORAL.2 system utilizes YTEX 0.8. MetaMap was called using the MetaMap UIMA annotator to allow for integration into our NLP framework which is also UIMA \cite{Ferrucci:2004kt} based. YTEX was run as a standalone system, and then a custom written UIMA annotator was used to transfer results from the YTEX database into a format compatible with our system. Hits from both systems were then processed in the same fashion, going through an identical set of annotators. Processing included filtering to remove high level stop concepts (20 in total) not typically used or useful for fine grained concept recognition. Two such examples are 'Disease' (C0012634) and 'Injury' (C0175677). Our system also removed concepts that had names containing text with 'M/mouse' and 'M/mice' as earlier work revealed some animal models of disease were being mapped to inappropriate UMLS semantic types. Finally we restricted resulting hits to those matching at least one of the requisite UMLS semantic types for this contest.  

Filtered hits generated by our system often failed to identify the full span of the annotation in some cases if the identified text started with acronyms and adjectives. For example, for concepts such as `LA enlargement' and `lower abdominal tenderness' present in the training data, our systems were able to capture `enlargement' and `abdominal tenderness' but left words `LA' and `abdominal tenderness' uncaptured. Thus, an additional post-processing step was done to improve results for the strict task. During the postprocessing step, we searched the concepts annotated with our models if they were preceded by any of the words `LV', `MCA',`LA',`abd',`PEA',`LE', `LGI', `ICA',`C2', `B12', `RCA', `RUQ', `GI', `VF', `lower', `chronic', but are not captured by our models. If so, the concepts were expanded to include these words within the concepts boundaries. These missing abbreviations were identified as the largest and most readily correctable error classes for both systems based on training data performance

It should be emphasized that the CORAL system framework built around both MetaMap and YTEX does not do any concept prediction itself, it simply refines predicted concept boundaries and removes concepts predicted by MetaMap and YTEX it determines are incorrect.

\subsection{Parameter Settings}
Default settings were used for YTEX, including a concept window of length 10 and the default INTRINSIC setting as a semantic similarity metric. MetaMap was run with the included word sense disambiguation server (-y option) and restricting allowable concepts to the SNOMED CT\textregistered and RXNORM vocabularies. This is implicit in the default YTEX configuration which only indexes these two vocabularies by default. The 2012AB distribution of UMLS was used by both programs. 

\subsection{Patient Tracking List (PTL) Data Set}
Prior to running data for the ShareClef task we evaluated both MetaMap and YTEX on PTL notes from the University of Alabama at Birmingham Health System. PTL documents consist of a summary of the patient's condition, with the majority of text in point form format along with some full sentences; a precise breakdown was not calculated. The document was irregularly formatted but highly structured, so in contrast to the ShareClef analysis, only disease mentions in the appropriate "problem" sections of the PTL document were analyzed. Document segmentation was done by a complex manually derived regular expression to identify the start and stop of the problem section. The same such expression was used for both YTEX and MetaMap. In total, there are 68 such annotated PTL documents with 603 annotated entities of which 223 are problems. Annotation was performed by two annotators (including one physician) and observed agreement was 91.9\% (uncorrected Cohen's kappa) \cite{carletta1996assessing}. 

\section{Results}
\subsection{Concept Boundary Detection Results}

Results for concept boundary detection (Task 1a) are shown in Table 1 and Table 2 for training and test data respectively.

\begin{table}
\caption{System Boundary Detection Results on Training Data}
\begin{center}
\begin{tabular}{| l | l | l | l | l | }
\hline
System & Task & Precision & Recall & Score  \\ \hline
TeamCORAL.1 (YTEX) & Strict & 0.512   & 0.440   & 0.473  \\ \hline
TeamCORAL.2 (MetaMap) & Strict & 0.722   & 0.460   & 0.562   \\ \hline
TeamCORAL.1 (YTEX) & Relaxed  & 0.915  & 0.639 &  0.752  \\ \hline
TeamCORAL.2 (MetaMap) & Relaxed & 0.875  & 0.556  & 0.680   \\ \hline
\end{tabular}
\end{center}
\end{table}

\begin{table}
\caption{System Boundary Detection Results on Test Data}
\begin{center}
\begin{tabular}{| l | l | l | l | l | }
\hline
System & Task & Precision & Recall & Score  \\ \hline
TeamCORAL.1 (YTEX) & Strict & 0.584   & 0.446   & 0.505  \\ \hline
TeamCORAL.2 (MetaMap) & Strict & 0.796   & 0.487   & 0.604   \\ \hline
TeamCORAL.1 (YTEX) & Relaxed  & 0.942  & 0.601 &  0.734  \\ \hline
TeamCORAL.2 (MetaMap) & Relaxed & 0.909  & 0.554  & 0.688   \\ \hline
\end{tabular}
\end{center}
\end{table}

Boundary detection was difficult for both systems. This was in part because neither MetaMap nor YTEX have the ability to annotate discontinuous concept boundaries limiting the effectiveness of both systems and effectively capping the maximum performance. Additionally MetaMap tended to include additional text (mostly prepositions and modifiers) that the ShareClef annotators did not. YTEX precision was significantly reduced by the inclusion of simple nouns when a compound noun was expected by the annotators. As a result YTEX performed poorly relative to MetaMap on the strict task.

Results were harder to compare for Task 1b, the mapping of text to CUIs. The PTL results (shown in Table 3) do not include accuracy data, since false negatives or false negative frames are not annotated in the PTL data set as they are in the ShareClef data set (Table 4). Instead results for precision, recall and F Score are shown for the PTL document.

\begin{table}
\caption{CUI Mapping Results on PTL Datasets}
\begin{center}
\begin{tabular}{| l | l | l | l | l | }
\hline
System & Precision & Recall & Score & Accuracy  \\ \hline
TeamCORAL.1 (YTEX)  & 55.72 & 68.33 & 61.38 &  NA \\ \hline
TeamCORAL.2 (MetaMap)  & 80.28 & 79.19 & 79.73 &  NA \\ \hline
\end{tabular}
\end{center}
\end{table}

\begin{table}
\caption{CUI Mapping Results on Training Data}
\begin{center}
\begin{tabular}{| l | l | l | }
\hline
System & Task & Accuracy   \\ \hline
TeamCORAL.1 (YTEX) & Strict & 0.414   \\ \hline
TeamCORAL.2 (MetaMap) & Strict & 0.422   \\ \hline
TeamCORAL.1 (YTEX) & Relaxed  & 0.939  \\ \hline
TeamCORAL.2 (MetaMap) & Relaxed & 0.916   \\ \hline
\end{tabular}
\end{center}
\end{table}

\section{Discussion}

Strict boundary was difficult for both systems, a varied set of error classes were generated. The biggest problem for strict boundary detection however was the exclusion of adjacent relevant text (modifiers), a deficiency that partly overcome by our boundary extension rules. Also abbreviations were particularly and consistently difficult,  both systems labelled text such as  'BACTERIA OCC' as Osteochondritis dissecans. In general, the MetaMap based CORAL.1 performed better than the YTex based CORAL.2 in the strict task. 

The relaxed task showed significantly higher scores for both systems, but both systems still failed to identify problems - particularly phrases containing common polysemous words such as "inability", as in "inability to walk'. Overall YTEX performed slightly better at this relaxed task than MetaMap, due to the inclusion of partially mapped annotations as fully scoring (YTEX had many such annotations) and the superior ability of YTEX to correctly identify polysemous text. 

In Task 1b a formatting error prevented our results from being processed but we show results for the training data and the PTL documents here. Results differed only slightly for YTEX and MetaMap. Common sources of error in both systems were stemmed from unrecognized abbreviations and low frequency concepts that neither the semantic distributional approaches used by YTEX or the heuristics and word sense disambiguation server employed by MetaMap could overcome. For example YTEX identifies the physician abbreviation "Dr." as diabetic retinopathy and MetaMap identified the word call in "Call or return immediately" as "c-ALL", a precursor B-cell lymphoblastic leukemia. Another small class of errors may be due to problems in the ShareClef annotation. For example the ShareClef annotation identifies fever as CUI-less instead of pyrexia, a synonym for fever. 

Results for both systems were significantly worse on the PTL data set (Table 3) than on the ShareClef data set (Table 4). The difference can be explained in large part to the annotation of the PTL data set, where the annotation guidelines specify that only the most precise concept possible should be annotated and thus penalizes YTEX which generates a larger number of more general (false positive) concepts . As described in the methodology section the PTL document set is an order of magnitude smaller than the ShareClef data set and thus less reliable. Nonetheless it underlies the fact that small differences in annotation guidelines can have a large impact on clinical information extraction evaluation.

One system that was also given serious consideration for a more formal evaluation was the NCBO annotator. However due to the difficulty of the VM setup (including the loading of source vocabularies) and concerns about being banned for sending thousands of queries to the NCBO web service we declined to investigate this option further.

Finally, a higher performance for YTEX could have likely been gained by parameter tuning. An earlier evaluation of YTEX sense disambiguation \cite{garla2012knowledge} revealed that no single semantic similarity performed best on all datasets and that parameter could have been adjusted to the training set. Additionally a higher context windows size for YTEX gave higher performance (at the cost of run time) and could have been adjusted upward to improve our performance here. MetaMap performance could have also been improved by taking advantage of the scoring information returned (not reported by YTEX) to select a more optimal cutoff level. 

In conclusion, given a choice between YTEX and MetaMap our results suggest that YTEX would be a better system for "off the shelf" concept mapping. Other factors such as active development and ability to scale favor YTEX.  However MetaMap may be a better choice for precisely identifying concept boundaries.

\section{Acknowledgements}

This project was supported by the UAB Center for Clinical and Translational Science - grant number UL1 RR025777 from the NIH National Center for Research Resources, and the UAB Office of the Vice President for Information Technology.

%
%

\begin{thebibliography}{9}
%

\bibitem{Aronson:2001jv}
Aronson, A.R.:
MetaMap Candidate Retrieval.
Technical Report, Lister Hill National Center for Biomedical Communications, National Library of Medicine (2001)

\bibitem{Aronson:2010kv}
Aronson, A.R. and Lang, F.M.:
An overview of MetaMap: historical perspective and recent advances.
J. Am. Med. Inform. Ass. 17 229--236 (2010)

\bibitem{banerjee2002adapted}
Banerjee, S. and Pedersen, T.:
An adapted Lesk algorithm for word sense disambiguation using WordNet.
Computational linguistics and intelligent text processing. 136--145 (2002)

\bibitem{carletta1996assessing}
Carletta, J.:
Assessing agreement on classification tasks: the kappa statistic.
Comp. Ling. 22 249--254 (1996)

\bibitem{Ferrucci:2004kt}
Ferrucci, D. and Lally, A.:
UIMA: an architectural approach to unstructured information processing in the corporate research environment.
Nat. Lang. Eng. 10 327--348 (2004)

\bibitem{garla:2011ml}
Garla, V. and Re III, V.L. and Dorey-Stein, Z. and Kidwai, F. and Scotch, M. and Womack, J. and Justice, A. and Brandt, C.:
The Yale cTAKES extensions for document classification: architecture and application.
J. Am. Med. Inform. Ass. 18 614--620 (2011)

\bibitem{garla2012knowledge}
Garla, V.N. and Brandt, C.:
Knowledge-based biomedical word sense disambiguation: an evaluation and application to clinical document classification.
2012 IEEE Second International Conference on Healthcare Informatics, Imaging and Systems Biology (HISB) 22--22 (2012)

\bibitem{Savova:2010hy}
Savova, G.K. and Masanz, J.J. and Ogren, P.V. and Zheng, J. and Sohn, S. and Kipper-Schuler, K.C. and Chute, C.G.:
Mayo clinical Text Analysis and Knowledge Extraction System (cTAKES): architecture, component evaluation and applications.
J. Am. Med. Inform. Ass. 17 507--513 (2010)

\bibitem{Suominen:h}
Suominen H, Salantera S, Velupillai S et al.:
Three Shared Tasks on Clinical Natural Language Processing.
Proceedings of CLEF 2013 To appear. (2013)

\end{thebibliography}

%

\end{document}